\documentclass[aps,prl,twocolumn,showpacs]{revtex4}
\usepackage{amssymb}
\usepackage{amsfonts}
\usepackage{graphicx}
\begin{document}


\title{Evidence for Two Gaps and Breakdown of the Uemura Plot in Ba$_{0.6}$K$_{0.4}$Fe$_2$As$_2$ Single
Crystals}

\author{Cong Ren$^*$, Zhao-sheng Wang, Hui-qian Luo, Huan Yang, Lei Shan, and Hai-Hu Wen$^{\star}$}
\affiliation{National Laboratory for Superconductivity, Institute of
Physics and Beijing National Laboratory for Condensed Matter
Physics, Chinese Academy of Sciences, P.O. Box 603, Beijing 100190,
China}

\begin{abstract}
We report a detailed investigation on the lower critical field
$H_{c1}$ of the superconducting Ba$_{0.6}$K$_{0.4}$Fe$_2$As$_2$
(FeAs-122) single crystals.  A pronounced kink is observed on the
$H_{c1}(T)$ curve, which is attributed to the existence of two
superconducting gaps. By fitting the data $H_{c1}(T)$ to the two-gap
BCS model in full temperature region, a small gap of
$\Delta_a(0)=2.0\pm 0.3$ meV and a large gap of $\Delta_b(0)=8.9\pm
0.4$ meV are obtained. The in-plane penetration depth
$\lambda_{ab}(0)$ is estimated to be 105 nm corresponding to a
rather large superfluid density, which points to the breakdown of
the Uemura plot in FeAs-122 superconductors.
\end{abstract}

\pacs{74.20.Rp, 74.25.Ha, 74.70.Dd}

\maketitle

\newpage

One of the crucial issues in understanding the superconducting
mechanism in recently discovered FeAs-based layered superconductors
\cite{a1,a2} is the pairing symmetry of the superconducting gap and
the nature of the low energy excitations. For LnFeAsO$_{1-x}$F$_x$
(FeAs-1111) superconductors with $x=0.04\sim 0.2$ and
Ln=La,Ce,Nd,Sm, reports on pairing symmetry are divided into two
categories: those favoring a gap with \cite{a3,a4,a5,a6,a7} or
without \cite{a8,a9,a10} nodes. In other aspect, the upper critical
field measurement \cite{a11}, tunneling spectroscopy \cite{a12}, NMR
\cite{a13}, magnetic penetration depth measurement \cite{a14,a15}
and M\"{o}ssbauer \cite{a16} experiments indicate the existence of a
two-gap nature. However, because most of the current experiments
were performed on polycrystalline samples, the experimental results
within the context of pairing symmetry have not yet reached a
consensus.

The emergence of (Ba,Sr)$_{1-x}$K$_x$Fe$_2$As$_2$ (FeAs-122)
\cite{a17,a18} hole-doped superconductors with transition
temperature $T_c$ up to 38 K has enriched the research in this area.
Although stoichiometric (undoped) BaFe$_2$As$_2$ shares the same
feature with undoped LnFeAsO of a spin density wave type magnetic
order and a structural transition at similar temperatures
\cite{a19,a20}, however, there are also some significant differences
between the properties of FeAs-1111 and FeAs-122 superconductors.
Most notable is the fact that FeAs-1111 superconductors seem to have
very low charge carriers density \cite{a2} and hence low superfluid
density \cite{a5,a21}. This may give explanation to the fact that
the Uemura plot is satisfied in the FeAs-1111 systems
\cite{a21,a22}. According to Uemura \emph{et.al.} \cite{a23}, in
superconductors with low superfluid density ($\rho_s$), $T_c$ scales
linearly with $\rho_s \propto \lambda^{-2}_{ab}(0)$, this is
actually not required by the BCS theory. However, it is found that
the charge carrier density in FeAs-122 is an order of magnitude
larger than those in FeAs-1111 systems \cite{a17,a24}. Thus it is
very interesting to know whether the Uemura plot is still satisfied
in FeAs-122 superconductors.

Lower critical field $H_{c1}(T)$, or equivalently, magnetic
penetration depth $\lambda(T)$ are fundamental probes of the nature
of the pairing symmetry, or/and multigap of unconventional
superconductors.  As an advantage, $H_{c1}(T)$ measurement probes
relatively large depths (in scale of $\lambda \sim 100$ nm) and are
insensitive to sample surface condition. In this Letter we present
the first detailed magnetic penetration measurements of
superconducting Ba$_{0.6}$K$_{0.4}$Fe$_2$As$_2$ single crystals. The
local magnetization measurements allow a precise determination of
$H_{c1}$. We found the presence of possibly a full gap feature
together with two gaps in Ba$_{0.6}$K$_{0.4}$Fe$_2$As$_2$
superconductors. Meanwhile, the absolute value of $H_{c1}(0)$
determined from this work place the samples far away from the Uemura
plot.

Crystals of Ba$_{0.6}$K$_{0.4}$Fe$_2$As$_2$ were grown by FeAs flux
method. The details of crystal synthesis have been described
elsewhere \cite{a24}. Our crystals were characterized by both
resistivity and ac susceptibility measurements with $T_c=36.2$ K and
a transition width of $\Delta T_c=0.45$ K (10\%-90\% of normal state
resistivity) \cite{a24}, as displayed in left inset of Fig.1 (a). A
single crystal (sample No. 1) was selected from the cleaved as-grown
bulk under optical microscope.  The sample has dimensions of 110
$\mu$m in diameter and 40 $\mu$m in thickness, as shown in the right
inset of Fig. 1(a). The crystal structure was examined by X-ray
Diffraction (XRD), and a typical diffraction pattern was shown in
the main panel of Fig. 1(a). Only (00$l$) peaks were observed with
the full-width-at-half-maximum (FWHM) around 0.1 $^{\circ}$, which
indicates good crystallization of our samples.

The local magnetization measurement was performed on two crystals
using a two dimensional electron gas (2DEG)-based micro Hall sensor
with an active area of $10\times 10\ \mu$m$^2$. The Hall sensor was
characterized without sample attachment at different temperatures,
the Hall coefficient $R_{H}$ was identical to 0.22 $\Omega$/Oe,
independent of $T$ below 80 K. In our experiment, we used a small
field sweep rate of 30 Oe/min to measure the isothermal
magnetization $M(H)$ curves in both decreasing ($M_{dec}$) and
increasing ($M_{inc}$) the fields to minimize the complex effects of
the character of the field penetration in a layered structure
\cite{a25}. A second crystal (sample No. 2) with dimensions of
$210\times 150\times 50$ $\mu$m$^3$ was also measured by Hall sensor
and vibrating sample magnetometry (VSM), and both measurements
showed essentially identical behavior except for the different
demagnetization effect.

In order to check whether the vortex entry is influenced by the
Bean-Levingston surface barrier\cite{a26,a27}, we measured the
magnetization hysteresis loops. The result is shown in the inset of
Fig. 1(b). The $M(H)$ curve shows a symmetric feature at $T$ = 32.1
K which is close to $T_c$, indicative of the dominance of bulk
pinning and the absence of the Bean-Levingston surface barrier for
vortex entry.

\begin{figure}
\includegraphics[scale=0.78]{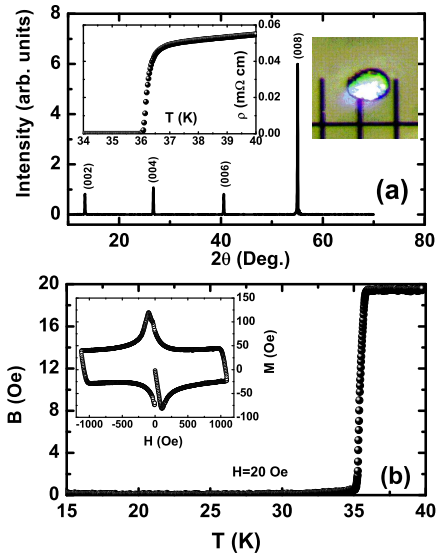}
\caption{\label{fig:fig1}(Color online) (a) X-ray diffraction
pattern of the Ba$_{0.6}$K$_{0.4}$Fe$_2$As$_2$ crystal. The right
inset shows the sample (No. 1) under the optical microscope. The
scale underneath with 100 $\mu$m/grid is used to display the size of
the sample. The left inset shows the resistive transition on one
crystal cut from the same piece as the one for the measurement of
$H_{c1}$. (b) The induction $B$ by Hall probe as a function of $T$
measured in zero-field-cooled mode with applied field. Inset: A
typical magnetization hysteresis loop measured at 32.1 K by the Hall
sensor.}
\end{figure}

The quality of the crystal and the accuracy of the local
magnetization measurement were demonstrated in Fig. 1(b), in which
the induction $B=H + 4\pi M$ sensed by the Hall probe is plotted as
a function of $T$ under a bias field $H=20$ Oe. The superconducting
transition can be detected as the induction jump at $T_c=35.8$ K
with a width of $\Delta T_c=0.5$ K, which shows the high quality of
the crystal. As shown in raw data of $B(T)$, at low temperatures,
$B$ is close to 0, implying a full Meissner shielding effect: $4\pi
M\simeq -H$. Alternatively, the magnitude of the change in $B$
measured by the Hall sensor through the superconducting transition
is 19.8 Oe, very close to the applied magnetic field 20 Oe. Thus the
achievement of full Meissner shielding effect in our measurement
provides a reliable way to determine the value of $H_{c1}$.

\begin{figure}
\includegraphics[scale=0.8]{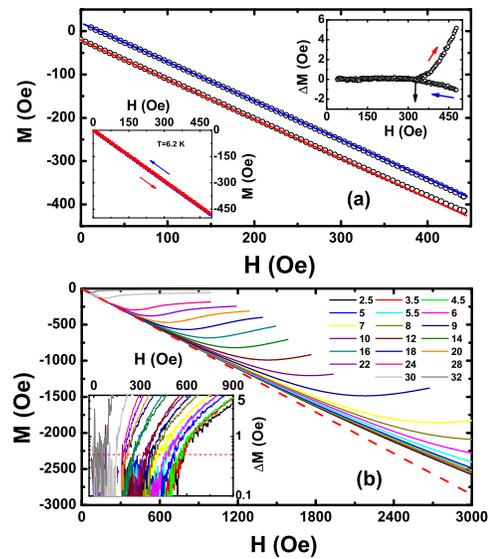}
\caption{\label{fig:fig2}(Color online) (a) A typical magnetization
hysteresis loop (symbols), the solid lines are the linear fitting
curves using the low field data (Meissner line). The increasing and
decreasing branches are shifted downward and upward, respectively
for clarity. The same case is shown in the bottom-left inset when
the maximum field is less than $H_{c1}$. The upper-right inset shows
the magnetization data subtracted the Meissner line. The arrows
indicate the direction of sweeping fields and the determination of
$H_{c1}$. (b) The initial part of the magnetization curves $M(H)$ of
sample No. 2 at various temperatures. The dashed line gives the
Meissner linear approach. Inset: The same magnetization data in
field subtracted by the Meissner line. The dashed line in the inset
sets up a criterion of 0.5 Oe. }
\end{figure}

Shown in Fig.2(a) are the typical isothermal $M(H)$ curves by taking
$M_{dec}$ and $M_{inc}$ at $T$=17.1 K, respectively. It can be seen
that, at low $H$, the $M_{dec}$ and $M_{inc}$ are fully reversible,
showing a common linear dependence of the magnetization on field as
displayed in the bottom-left inset of Fig. 2(a).  At high $H$, a
deviation from the linear dependence occurs at $H=H_{c1}$ for both
$M_{dec}$ and $M_{inc}$ curves. To distinguish these deviations, we
fit more than 50 data points between 10 and 30 Oe by a linear
relation to account for the common linear dependence of $M(H)$.
These fitted linear lines describe the Meissner shielding effects
(``Meissner line'') at low fields, as evidenced quantitatively in
Fig. 2(a) in which the slope of the fitted lines are -0.98, being
very close to -1. Thus the deviation of $M(H)$ from the linear
Meissner shielding is an indication of the first penetration field
$H_{c1}$. An alternative way to determine the value of $H_{c1}$ from
these reversible isothermal $M(H)$ curves is to subtract the
Meissner line from both $M_{dec}(H)$ and $M_{inc}(H)$ curves, as
illustrated in the upper-right inset of Fig. 2(a). The threshold
field of non-zero magnetization happens to be the divergence of the
increasing and decreasing $M(H)$ curves. It is noted that the values
of $H_{c1}$ with this criterion were determined in both increasing
and decreasing field, so they are the true thermodynamic values and
are not altered by the surface barrier \cite{a25}.

For a strict treatment, we determined the value of $H_{c1}$ by
examining the point of departure from the Meissner line on the
initial slope of the $M(H)$ curve. In the inset of Fig. 2(b) we show
how to determine $H_{c1}$ by using a criterion of $B=\Delta M =0.5$
Oe at different temperatures. For a quantum flux $\phi_0=20.7$ Oe
$\mu$m$^2$, $\Delta M =0.5$ Oe is equivalent to about $(2\sim 3)
\phi_0$ penetrating into the $(10\times 10)\ \mu$m$^2$ sensing area,
which is the limit of our Hall probe technique. The $H_{c1}$ values
determined in this way are about 4\% larger than those estimated
from the point where the reversible magnetization deviates from
linearity, and we did not observe any significant difference in the
$T$-dependence of $H_{c1}$ deduced from either of the two criterion.

Shown in Fig. 3(a) and (b) are the main results of our experiment,
in which the obtained $H_{c1}$ are plotted as a function of $T$ for
crystals No. 1 and 2, respectively.  At $T<4\sim 5$ K, $H_{c1}(T)$
is weakly $T$-dependent and seems to show a tendency towards
saturation at lower $T$ (in the limited temperature range). As
illustrated in the insets of Fig. 3(a) and (b), the saturated
$H_{c1}$ reach 695 Oe for sample No. 1 and 590 Oe for sample No. 2.
This tendency of $H_{c1}(T)$ reflects a possible fully gapped nature
of superconducting state at low $T$ for
Ba$_{0.6}$K$_{0.4}$Fe$_2$As$_2$ superconductors, although we could
not rule out the possibility of a small gap with nodes in the dirty
limit.

A pronounced \emph{kink} can be easily observed in $H_{c1}(T)$
curves at $T \sim 15$ K for both samples. Obviously, the occurrence
of the kink in $H_{c1}(T)$ can not be explained by the model with an
$s$-wave or $d$-wave \emph{single gap}.  On the other hand, this
kinky structure in $H_{c1}(T)$ resembles that of the related
penetration depth of the two-band superconductor MgB$_2$ \cite{a28},
in which a positive curvature was observed and explained by the
multi-band theory \cite{a29}. In addition, recent ARPES measurement
resolved a two-gap nature in a similar
Ba$_{0.6}$K$_{0.4}$Fe$_2$As$_2$ crystal \cite{a30,a31}. Thus our
observation of a kink in $H_{c1}(T)$ strongly suggests the existence
of multiple gaps in Ba$_{0.6}$K$_{0.4}$Fe$_2$As$_2$ superconductors,
being consistent with that predicted in electronic band structure
calculations \cite{a32}.

\begin{figure}
\includegraphics[scale=0.85]{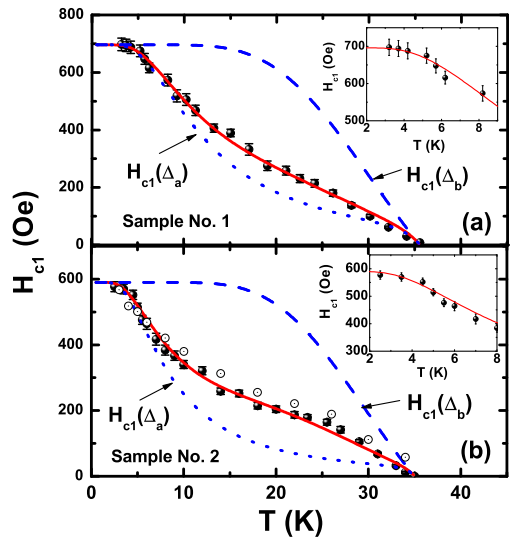}
\caption{\label{fig:fig4}(Color online) The extracted $H_{c1}$ as a
function of $T$ for sample No.1 (a) and No.2 (b)(full circles with
error bars). The open dotted circles in (b) represent the $H_{c1}$
determined from the VSM measurements taken into account of the
demagnetization effect. The solid lines are the fitting curves using
the two-gap model [Eq.(1) and Eq.(2)]. The contributions of the
small gap $[H_{c1}(\Delta_a)]$ and the large gap
$[H_{c1}(\Delta_b)]$ in the model are also shown by the dotted and
dashed lines, respectively. The two insets in (a) and (b) show the
enlarged views of the data $H_{c1}(T)$ in low temperature region
together with the theoretical fitting curves (solid lines) with a
two-gap model.}
\end{figure}

Quantitatively we analyze and discuss in more detail the data of
$H_{c1}(T)$ to elucidate the related gap function. For our crystals,
assuming $\lambda_{ab}(0)\sim 100-200$ nm \cite{a33}, the coherence
length $\xi_{ab}(0)$ was estimated to be 2-2.5 nm from an extremely
high upper critical field $H_{c2}$
[$\mu_0H_{c2}^{//c}(0)=\phi_0/2\pi \xi^2_{ab}(0)$] \cite{a24,a34},
and the mean free path (determined from the resistivity at 38 K) is
$\sim 15$ nm, our samples are therefore expected to be in the
moderately clean, local limit. In this case the local London model
is valid to describe the data. For a single gap superconductor,
$H_{c1}$ relates the normalized superfluid density as:
$\widetilde{\rho_s}(T)\equiv
\lambda_{ab}^2(0)/\lambda_{ab}^2(T)=H_{c1}(T)/H_{c1}(0)$, and
$\widetilde{\rho_s}(T)$ is given by \cite{a35,a36}
\begin{equation}
\widetilde{\rho_s}(T)=1 + 2
\int_{0}^{\infty}\frac{\textrm{d}f(E)}{\textrm{d}
E}\frac{E}{\sqrt{E^2-\Delta(T)^2}}\textrm{d}E
\end{equation}
with $f$ the Fermi function. Here the total energy is
$E=\sqrt{\epsilon^2 + \Delta^2}$, and $\epsilon$ is the
single-particle energy measured from the Fermi surface. It is
assumed that the gap $\Delta$ on each Fermi surface follows the
weak-coupling BCS temperature dependence. For a superconductor with
two gaps, the normalized superfluid density may be written as
\begin{equation}
\widetilde{\rho_s}=x \widetilde{\rho_s}^a +
(1-x)\widetilde{\rho_s}^b,
\end{equation}
where $x$ is the fraction of superfluid density
$\widetilde{\rho_s}^a$ associated with the small gap $\Delta_a$. The
results of the calculations and $H_{c1}$ of sample No. 1 and 2 are
shown by the red solid lines in Fig. 3(a) and (b), respectively.
Fitting the data to above equations (with two $s$-wave gaps) yields:
$\Delta_a=1.6\pm 0.3$ meV, $\Delta_b=9.1\pm 0.3$ meV and $x$=0.72
for sample No.1, and $\Delta_a=2.2\pm 0.2$ meV, $\Delta_b=8.8\pm
0.3$ meV and $x$=0.70 for sample No.2. The gaps obtained from our
$H_{c1}(T)$ measurements are clearly smaller than those determined
from the ARPES measurements\cite{a30,a31}. This discrepancy may be
induced by the different ways and different criterions in
determining the gaps, this should be checked by future experiments.
It is interesting to note that the large gap accounts for only 30\%
of the total superfluid density. We must stress that although a
small gap with nodes (in the dirty limit) cannot be excluded from
our low temperature data, this will not lead to a significant change
to the general fitting results obtained here.

Another important result of our experiment, the absolute value of
$\lambda_{ab}(0)$ is \emph{independently} deduced from the measured
$H_{c1}(0)$. In principle, we evaluate the penetration depth
$\lambda_{ab}(0)$ using the expression:
$H_{c1}^{//c}=(\phi_0/4\pi\lambda^2_{ab})\ln \kappa$, where
$\kappa=\lambda/\xi$ is the Ginzburg-Landau parameter (here we
assume that $\kappa$ is $T$-independent). Using
$H_{c1}^{//c}(0)=695$ Oe and $\kappa=80$, we obtain
$\lambda_{ab}(0)\simeq 105$ nm for sample No. 1 and 115 nm for
sample No. 2. The value of $\lambda_{ab}^{-2}(0)$, or equivalently
the condensed carrier density $n_s/m^{*}$ (superconducting carrier
density/effective mass), allows us to check whether the well-known
scaling behaviors between $n_s/m^{*}$ and $T_c$ still works for the
present system. In Fig. 4 we present our results together with many
others, including the FeAs-1111 system, cuprates, MgB$_2$ and
NbSe$_2$ \cite{a37}. It is remarkable that
Ba$_{0.6}$K$_{0.4}$Fe$_2$As$_2$ resides far away from the Uemura
plot, which is contrasted by the case of the FeAs-1111 system which
is quite close to the Uemura plot. This discrepancy between FeAs-122
and FeAs-1111 implies very essential difference between the physics
in the two systems, this warrants further investigations.

\begin{figure}
\includegraphics[scale=0.9]{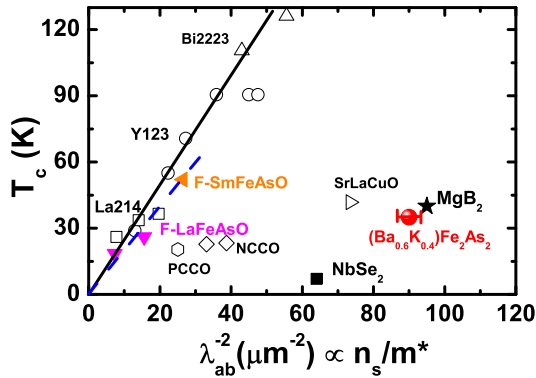}
\caption{\label{fig:fig4}(Color online) The correlations between
$T_c$ and the superfluid density $n_s/m^*$. Points for the cuprates
are taken from Ref.21,22. The full triangles are showing the data
for F-LaFeAsO (Ref.5 and Ref.21) and F-SmFeAsO (Ref.13 and our
work), both are close to the Uemura plot of the cuprate
superconductors. The data of the FeAs-122 phase (the red full
circle), MgB$_2$ and NbSe$_2$ are far away from the general Uemura
plot \cite{a23}.}
\end{figure}

To summarize, we conduct magnetization measurements on
Ba$_{0.6}$K$_{0.4}$Fe$_2$As$_2$ single crystals, and the lower
critical field $H_{c1}(T)$ is reliably extracted. It is found that
$H_{c1}$ exhibits a pronounced kink at $T\sim 15$ K, which indicates
a multi-gap nature. By using the two-gap weak coupling BCS model to
fit the data, we obtained a small gap of $\Delta_a(0)\simeq 2.0\pm
0.3$ meV and a large gap of $\Delta_b(0)\simeq 8.9\pm 0.4$ meV. An
estimate of the in-plane penetration depth gives
$\lambda_{ab}(0)\simeq 105$ nm, which points to the breakdown of the
Uemura relation for the optimally doped
Ba$_{0.6}$K$_0.4$Fe$_2$As$_2$ superconductors.

Acknowledgement: The authors are grateful to Junren Shi, Tao Xiang
and Jan Zaanan for helpful discussions, and Wen-Xin Wang and Hong
Chen for providing us the GaAs/AlGaAs substrates. This work is
supported by the Natural Science Foundation of China, the Ministry
of Science and Technology of China (973 project No: 2006CB60100,
2006CB921107, 2006CB921802), and Chinese Academy of Sciences
(Project ITSNEM).

$^*$ cong\_ren@aphy.iphy.ac.cn

$^{\star}$ hhwen@aphy.iphy.ac.cn

\end{document}